\documentclass[aps,twocolumn,prl,superscriptaddress,floatfix,preprintnumbers]{revtex4}

\usepackage{graphicx,bm,amsmath}

\font\cmss=cmss12 
\def\1{\hbox{{1}\kern-.25em\hbox{l}}}
\def\bfZ{\relax{\hbox{\cmss Z\kern-.4em Z}}}

\begin{document}


\title{Geometric scaling in inclusive ${\bf eA}$ reactions and 
nonlinear perturbative QCD}

\author{A.~Freund}
\affiliation{Institut f{\"u}r Theoretische Physik, 
Universit{\"a}t Regensburg, D-93040 Regensburg, Germany}
\author{K. Rummukainen}
\affiliation{NORDITA, DK-2100 Copenhagen, Denmark 
  \&  Department of Physics, P.O.Box 64, FIN-00014
   University of Helsinki, Finland
}
\author{H.~Weigert}
\affiliation{Institut f{\"u}r Theoretische Physik, 
Universit{\"a}t Regensburg, D-93040 Regensburg, Germany}
\author{A.~Sch{\"a}fer}
\affiliation{Institut f{\"u}r Theoretische Physik, 
Universit{\"a}t Regensburg, D-93040 Regensburg, Germany}
\date{\today}

\begin{abstract}
   In this note we report on geometric scaling in inclusive $eA$
   scattering data from the NMC and E665 experiments. We show that
   this scaling, as well as nuclear shadowing, is expected in the
   framework of nonlinear pQCD at small $x$ based on a simple
   rescaling argument for  $ep$ scattering.
\end{abstract}

\preprint{HIP-2002-48/TH, NORDITA-2002-68/HE}

\maketitle

\section{Introduction}

Perturbative QCD, in particular where based on factorization theorems for
high energy processes, has been very successful in describing data from
high energy experiments. However, perturbative evolution has almost
always been {\it linear} in the parton densities, which was, and very
often still is, quite sufficient to describe high energy scattering.

There are, however, situations where this is bound to fail. Among
those are high energy scatterings at energies large enough to create a
gluon medium in which multi--gluon correlations can no longer be
neglected. This has first been discussed in the context of deep
inelastic scattering of electrons off (large) nuclei, where both the
large number of nucleons and the high energy lead to an increase in
the number of gluons involved in the
scattering~\cite{McLerran:1994ni,McLerran:1994ka,Jalilian-Marian:1997xn,
  Balitsky:1996ub,Jalilian-Marian:1997gr,Jalilian-Marian:1997dw,
  Weigert:2000gi,Iancu:2000hn,Ferreiro:2001qy}.  In this regime
nonlinear effects, leading to a taming or saturation/unitarization of
$F_2^A$, should be included.

In the guise of a very successful phenomenological fit, this idea has
even been applied to the $e p$ experiment at HERA with its strong and,
if untamed, unitarity violating rise of the proton structure function
$F_2^p(x_{\text{bj}},Q^2)$ with decreasing
$x_{\text{bj}}$~\cite{Abt:1993cb,Derrick:1993ft}.  This has led to the
observation of ``geometric scaling'' satisfied by HERA data in the
small $x_{\text{bj}}$ region~\cite{Stasto:2000er}. Our main focus here
will be an attempt to extend this geometric scaling idea to nuclei and
to check if it is already confirmed by available $e A$ data.

The question of the onset of this saturation/unitarization in
inclusive QCD observables with increasing energy (decreasing
$x_{\mathrm{bj}}$), in particular in the proton structure function
$F_2^p(x_{\mathrm{bj}},Q^2)$, has been the subject of active
discussions for many years (see a detailed discussion of this subject
in \cite{Weigert:2000gi,Stasto:2000er} and references therein.).
Phenomenologically, the key feature at small $x_{\mathrm{bj}}$ is the
simple structure of the total $\gamma^*p$ cross section or $F_2^{p,A}$
that arises directly from the underlying physics picture.  Viewed in
the infinite momentum frame of the target, the virtual photon splits
into a $q\bar q$ color-dipole (treated as eikonalized Wilson-lines)
with a distribution characterized by $Q^2$.  This dipole then punches
through, and interacts with the target, which it sees as a pancake of
infinitesimal longitudinal thickness.  Consequently the cross section
appears as a convolution of the square of a photon wave function,
which gives the probability to create the $q\bar q$ pair, and a dipole
cross section, which contains all the information about the target and
the strong interaction physics.

The latter carries $x_\text{bj}$ dependence which arises from gluonic
fluctuations which induce an increase in the number of
gluons~\cite{freakfootnote}.
From this point of view, the
target will display a growing density of gluons per transverse area
(integrated over the longitudinal extent of the target) that drives
the system into a saturation regime.  The corresponding formula for
the total cross section in $\gamma^*p$ is then
\begin{eqnarray}
\label{eq:stotfact}
\sigma_{\mathrm{tot}}(x_{\mathrm{bj}},Q^2)& = & \int d^2{\bm z}\int\limits^1_0\!d\alpha\
|\psi_{\gamma^*}(\alpha,{\bm z}^2,Q^2)|^2
\nonumber\\& & \hspace{1.5cm}\times\
\sigma_{\mathrm{dipole}}(x_{\mathrm{bj}},{\bm z}^2)
\end{eqnarray}
where $\alpha$ is the longitudinal momentum fraction of the $q$ or $\Bar
q$, and $\bm{z}$ their relative transverse separation. Note that the
$Q^2$ and $x_{\mathrm{bj}}$-dependence is clearly separated into wave
function and dipole cross section. This reflects the natural
distinction between the transverse and longitudinal directions seen at
high energies.  $Q^2$ sets the scale for transverse resolution, but
longitudinal information about the target is subsumed in the
$x_\text{bj}$ dependence of the dipole cross section. The idea is that
the dipole cross section depends on $x_\text{bj}$ only via a scale
$Q_s(x_\text{bj})$ as $\sigma_{\mathrm{dipole}}(x_{\mathrm{bj}},{\bm
  z}^2)=\sigma_{\mathrm{dipole}}({\bm z}^2Q_s^2(x_\text{bj}))$.  $Q_s$ is
the saturation scale, i.\ e.\ the scale below which the gluonic
content of the target becomes ``black'' or $\sigma_\text{dipole}$ reaches
its large size asymptotics.

Together with Eq.~\eqref{eq:stotfact} this implies that
$\sigma_\text{tot}$ and hence $F^p_2$ scales according to
\begin{equation}
  \label{eq:scaling}
 \sigma_{\mathrm{tot}}(x_\text{bj},Q^2) =
 \sigma_{\mathrm{tot}}\Big(x_0,\frac{Q^2}{Q_s^2(x_\text{bj})}Q_0^2\Big)
\ . 
\end{equation}
This phenomenon is called geometric scaling and, as already mentioned,
was first observed to be present in HERA data~\cite{Stasto:2000er}.
The fit used in~\cite{Stasto:2000er} assumes an ``eikonal'' shape for
$\sigma_\text{dipole}$ that saturates for large dipoles. As
$x_\text{bj}$ decreases this saturation affects smaller and smaller
distances, inducing $Q_s$ to grow, a fact which was parametrized by
\begin{equation}
  \label{eq:Q_s}
  Q_s(x_\text{bj}):=\Big(\frac{x_0}{x_\text{bj}}\Big)^\lambda Q_0
\ .
\end{equation}

The theoretical basis for such saturation/unitarization features, with a
scaling behavior as in Eq.~\eqref{eq:scaling}, together with precisely
the functional dependence of $Q_s$ as in Eq.~\eqref{eq:Q_s}, lies in
nonlinear evolution equations that try to predict this $x_\text{bj}$
dependence from first principles.

The most promising approach to properly include
saturation/unitarization is the JIMWLK equation~\cite{Weigert:2000gi}
which resums the leading $\ln(1/x_{\text{bj}})$ terms in {\it all}
N-point correlation functions of the participating fields not just the
leading correlators as in $k_{\perp}$-factorization.  This leads to a
non-linear renormalization group equation (RGE) for the generating
functional for these N-point correlators which slows down the rapid
growth of structure functions at small $x_\text{bj}$ while at the same
time preventing the system from drifting into the infrared. This is
the basis for a theoretically sound, selfconsistent treatment.
Linearizing the JIMWLK equation, one recovers the well-known BFKL
equation (see \cite{Weigert:2000gi,Jalilian-Marian:1997jx} for
details) as its low density limit. We thus learn to interpret the loss
of infrared safety in BFKL as being due to the absence of any nonlinear
corrections and the predicted too rapid a growth in
$F_2^p(x_{\text{bj}},Q^2)$ as signals of BFKL having exceeded its
range of validity.

The full content of the JIMWLK equation can be accessed numerically:
being a functional equation of Fokker-Plank type it can be rewritten
as a Langevin equation with white noise.  This then can be discretized
and solved using lattice methods~\cite{HW-KR}. This yields the
$x_\text{bj}$ dependence of $\sigma_\text{dipole}$ and in particular
the evolution ``rate'' $\lambda$ of Eq.~\eqref{eq:Q_s}.

Of particular interest to us in this context is the fact that the
evolution equation is {\it target independent}. Target dependence
arises only from the initial conditions for the $x_{\text{bj}}$
dependence of the dipole cross section (related to the initial
distribution of gluons in the transverse plane), which will, in
particular, be $A$ dependent.  These facts will allow us to compare
$ep$ and $eA$ scattering and investigate whether one of the basic
predictions of this approach, geometric scaling, can also be found in
inclusive $eA$ processes.

\section{Geometric scaling and nuclear shadowing within nonlinear pQCD.}

Before talking about geometric scaling in $eA$ scattering, one should
address nuclear shadowing or the suppression of $\frac{1}{A}F^A_2$  to $F^p_2$ at
small $x_{\text{bj}}$. If this is at
all possible it increases our phenomenological leverage considerably
by allowing us to compare experiments with different $A$ and
$x_{\text{bj}}$ directly. This is of particular importance as the
available $eA$ data do not offer any leverarm in $x_{\text{bj}}$ for a
given $Q^2$ or vice versa.

In fact, the leading effect of nuclear shadowing can be easily
incorporated in this approach by realizing that the change in
transverse size simply changes the overall normalization of
$\sigma_\text{dipole}$ while the change in longitudinal extent can be
incorporated by increasing our initial $Q_s$ by a factor $A^{\delta}$.
In the simplest case, assuming the distribution of spectator partons
in the target to be homogeneous, $\delta = 1/3$.
Eq.~\eqref{eq:scaling} for $F_2 \sim \sigma \cdot Q^2$ then turns into
\begin{equation}
\label{eq:F2}
\left(\frac{x_{\text{bj}}}{x_0}\right)^{2\lambda}\frac{F^A_2(x_{\text{bj}},Q^2)}{A^{2/3+1/3}} = F^p_2\left(x_0,\left(\frac{x_{\text{bj}}}{x_0}\right)^{2\lambda}\frac{Q^2}{A^{1/3}}\right).
\end{equation}
Note that $0\leq x_{\text{bj}}\leq A$ since $x_{\text{bj}}$ is given
per nucleon. The power $2/3$ in the overall $A$ dependence stems from
a simple geometric factor, the area of the target, in the impact
parameter integral in $\sigma_{\text{dipole}}$. Thus we have an
unambiguous prediction for $\frac{1}{A}F^A_2$ which states that not
only can the observed nuclear shadowing in the $eA$ data be explained
by a simple rescaling of the variable $Q^2$ with $A^{1/3}$, but also
that all $\frac{1}{A}F^A_2$ data should, plotted vs.  $\tau=
\left(\frac{x_{\text{bj}}}{x_0}\right)^{2\lambda}\frac{Q^2}{A^{1/3}}$,
lie on the same curve as the data for $F^p_2$. Hence, this approach
also predicts geometric scaling in inclusive $eA$ scattering analogous
to the one observed in $ep$ scattering.

\section{Geometric scaling in $\frac{1}{A}F^A_2$}

Note that Eq.\ (\ref{eq:F2}) is strictly true only
in the asymptotic region of small $x_{\text{bj}}$, large $Q^2$ and large
nuclei.  Therefore, we expect that there might be nonperturbative
corrections which can spoil the form of Eq.\ (\ref{eq:F2}) and
consequently we perform a fit to the combined nuclear data with the
following form:
\begin{equation}
\label{eq:F2fit}
\left(\frac{x_{\text{bj}}}{x_0}\right)^{2\lambda}\frac{1}{A^{\gamma}}\frac{1}{A}F^A_2(x_{\text{bj}},Q^2) = F^p_2\left(x_0,\left(\frac{x_{\text{bj}}}{x_0}\right)^{2\lambda}\frac{Q^2}{A^{\delta}}\right),
\end{equation}
where $\gamma$ and $\delta$ are the fit parameters, which ideally would be $0$
and $1/3$ respectively.

\begin{figure}[tp]
\begin{center}
\includegraphics[height=7.92cm]{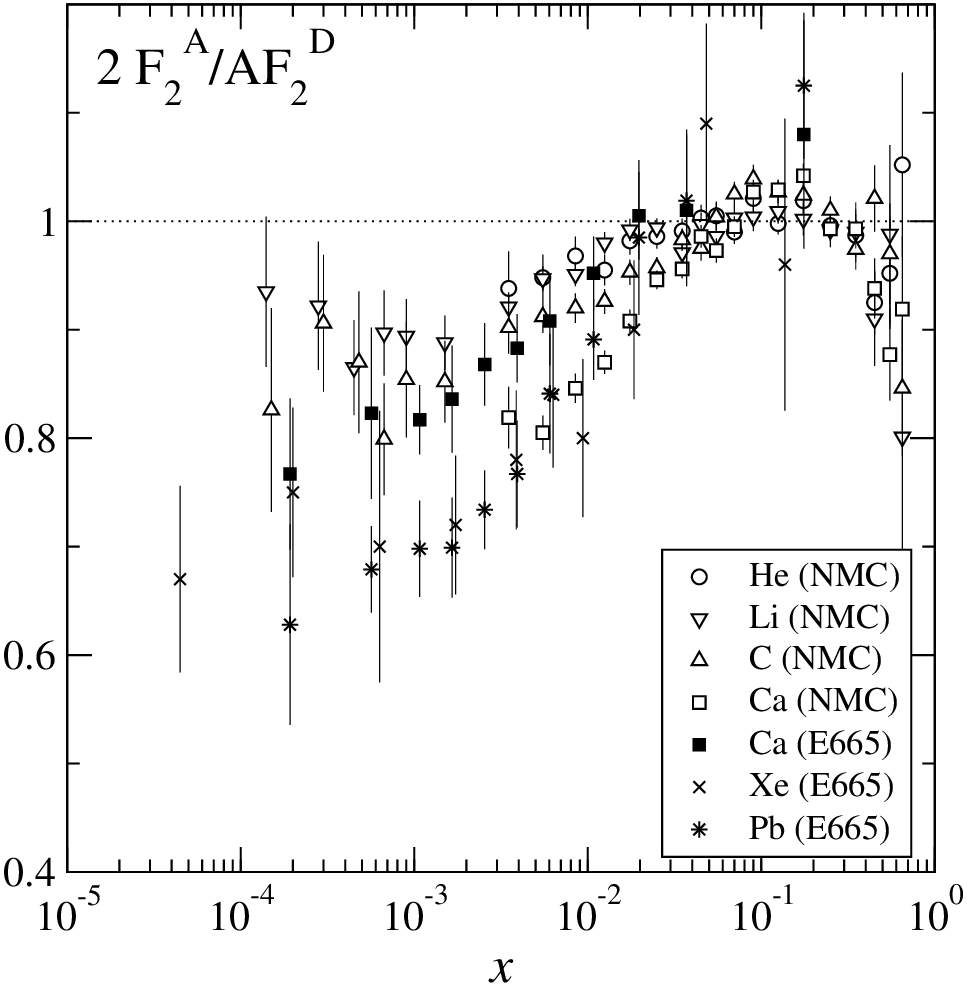}
\caption{Ratio of $2F_2^A/AF_2^D$ vs.\ $\tau$ for NMC and E665.
Note the increasing steepness with $A$  within the NMC and E665 
datasets respectively.}
\label{fig:rationuc}  
\end{center}
\end{figure}

The starting point of our analysis is the published
NMC~\cite{Arneodo:1997kd} and E665~\cite{Adams:1995ri,Adams:1995is}
data for the ratios of nuclear to deuterium structure functions
$2F_2^A/AF_2^D$.  In Fig.~\ref{fig:rationuc} we show the ratios as
functions of $x$.
From these ratios we obtain the nuclear structure functions $F_2^A$ as
follows: we first note that NMC data show the deuterium to
proton structure function ratio being $F_2^D/2F_2^p = 1$ to better than 95\%
accuracy in the dynamical range of interest ($x_{\text{bj}} <
0.1$)~\cite{Arneodo:1997kd}, which is well within the error margins of
the nuclear $F_2$ measurements.  Thus, we convert $F_2^A/F_2^D$ to
structure functions $F_2^A$ by multiplying the ratios by $F_2^p$,
using the geometric scaling ansatz to extrapolate $F_2^p$ to the
appropriate values of $x_{\text{bj}}$ and $Q^2$.

\begin{figure}[tp]
\begin{center}
\includegraphics[height=8cm]{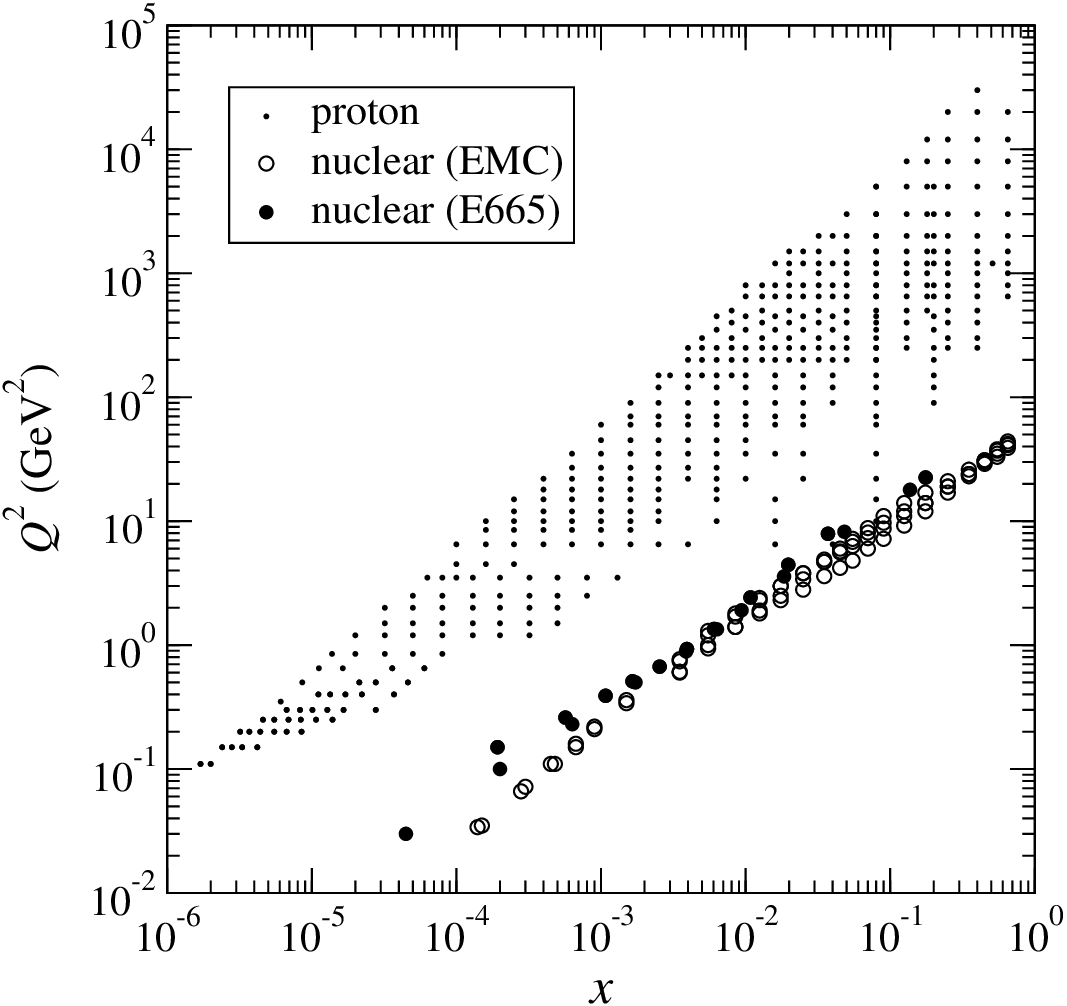}
\caption{Comparative phase space of HERA, NMC and E665.}
\label{fig:phase}  
\end{center}
\end{figure}

This rescaling is necessary since the kinematic ranges of the HERA,
NMC and E665 experiments are not identical as shown in
Fig.~\ref{fig:phase}.

Now we are ready to use Eq.~\eqref{eq:F2fit} to compare nuclear and
proton structure functions. In Fig.~\ref{fig:rationuc} the rescaled
$F_2^A$ is seen to fall on top of the dashed line representing the
geometric scaling fit to the HERA $F_2^p$. For comparison we also show
the quality of the scaling fit within the HERA data, offset by a
factor of 5.

The best fit to the combined nuclear data is achieved for $\gamma =
0.09$ and $\delta = 1/4$ which are close to but not quite the
asymptotic values the nonlinear pQCD approach predicts. The remaining
parameter $\lambda\sim 0.18$ in Eq.~\eqref{eq:F2fit} is already
determined from the geometric scaling fit to the HERA data.  With this
value for $\lambda$, the saturation scale at the upper end of the HERA
shadowing region at $x_0=10^{-1}$ comes out to be marginally
``perturbative'', $Q_s^2=1~\text{GeV}^2$.

\begin{figure}[tp]
\begin{center}
\includegraphics[height=8cm]{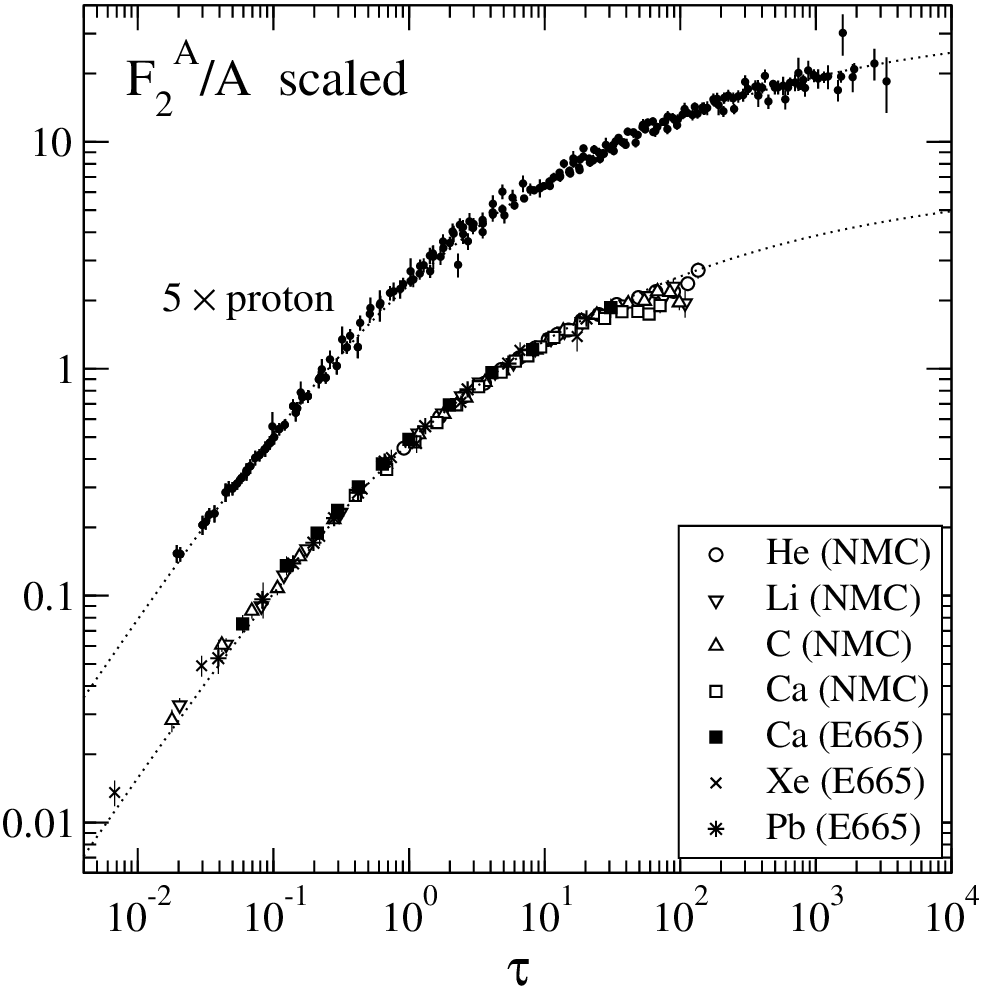}
\caption{Scaling behavior of NMC and E665 $F^A_2$ data vs. 
  $\tau =
  \left(\frac{x_{\text{bj}}}{x_0}\right)^{2\lambda}\frac{Q^2}{A^{1/3}}$.
  The vertical axis corresponds to the l.h.s. of Eq.~\eqref{eq:F2fit}.
  The dashed line corresponds to the geometric scaling curve obtained
  from HERA data. These are shown offset by a factor of 5.}
\label{fig:scaling}
\end{center}
\end{figure}

\begin{figure}[htbp]
\begin{center}
\includegraphics[height=8cm]{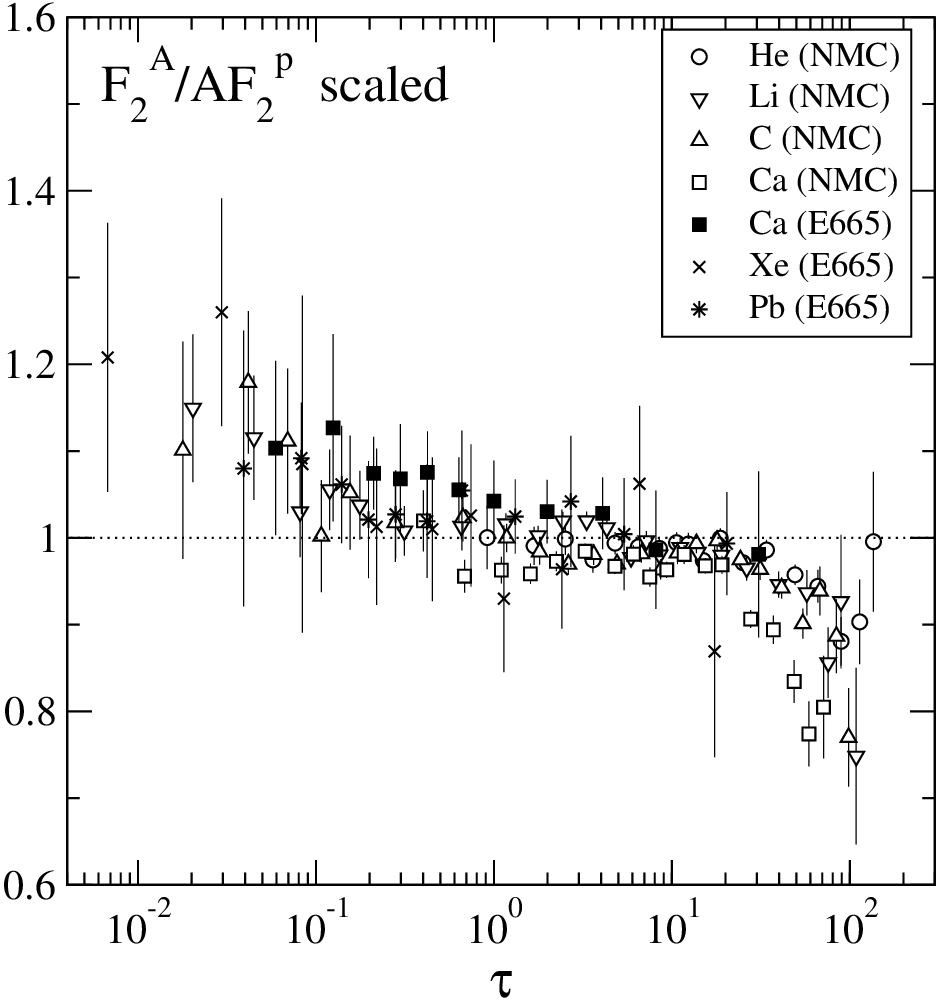}
\caption{Plot  of scaled NMC and E665 $F^A_2/AF_2^p$ data vs. 
  $\tau =
  \left(\frac{x_{\text{bj}}}{x_0}\right)^{2\lambda}\frac{Q^2}{A^{1/3}}$.
  This corresponds to ``dividing the data shown in
  Fig.~\ref{fig:scaling} by the dashed line.''}
\label{fig:scaling-ratios}
\end{center}
\end{figure}

We found that the NMC calcium data allows a smaller value of $\gamma$
and, more importantly, a value of $\delta = 1/3$ if the data on
lighter nuclei is left out. This can be seen best in a log-linear plot
of the scaled $F_2$ ratios shown in Fig.~\ref{fig:scaling-ratios},
where the plateau from $\tau=0.1$ to $\tau=10$ shows good overall
scaling behavior. $\delta=1/3$ would lift the NMC Ca data fully up to
1. This seems to indicate that the Ca nucleus is large enough for
the basic assumptions of the nonlinear pQCD approach to be valid. A
very encouraging sign, indeed.

Also note that there is a consistent difference in the E665 and NMC
data with the E665 data being much flatter in the ratio
$2 F_2^A/A F_2^D$, as evidenced in the comparison between the two
calcium data sets and shown in Fig.~\ref{fig:scaling-ratios}.  This
apparent inconsistency between the two data sets disappears when the
ratio $12 F_2^A/A F_2^C$ is compared between the two experiments
\cite{nmc1}. This question needs to be addressed experimentally.

Notwithstanding all of the above, nuclear data from the proposed
Electron Ion Collider at BNL on heavy nuclei at small $x_{\text{bj}}$
and large $Q^2$ is necessary to be absolutely definitive about
geometric scaling in inclusive $eA$ scattering.  These experiments
will by design provide data at smaller $x_{\text{bj}}$ but one should
take care to open up the phase space region.  Only if we have access to
a range of $x_{\text{bj}}$ values at any given $Q^2$ (and vice
versa) for each species $A$, can we obtain independent fits for the
evolution rate $\lambda$. This would be a prerequisite to
experimentally disentangle and confirm the $A$ and $x$ dependence
extracted simultaneously above. As the errors on the nuclear
data get smaller one would also require a direct measurement of
$F_2^A/AF_2^p$ that would allow us to do away with our rather cavalier
treatment of $F_2^D/2F_2^p$.

\section{Conclusions}

To summarize, we demonstrate for the first time that the NMC and E665
data for the nuclear structure function $\frac{1}{A}F^A_2$ also
exhibit, modulo the caveats given above, geometric scaling as already
seen in the nucleon structure function $F^p_2$. This has been
confirmed using the shadowing features of the nonlinear pQCD approach,
which, as the dominant effect, consists of a simple rescaling of
$Q^2_s$ by $A^{1/3}$. With the presently available nuclear data which
are at rather low $Q^2$ we had to allow slight modifications in the
powers entering the the $A$ dependence in order to get a perfect fit.
The NMC data for calcium, the heaviest nucleus available within this
dataset, would in fact be compatible with the $A^{1/3}$ scaling.
Given the limited phase space regions covered in the data sets we find
it very encouraging how well the geometric scaling in $eA$ follows the
one in $ep$.

AF was supported by the Emmi-Noether Grant of the DFG, AS by
BMBF and HW by the DFG Habilitanden-program.


\begin{thebibliography}{18}
\bibitem{freakfootnote}
\bibinfo{howpublished}{This clean separation is in fact a scheme choice:
  Depending on the frame these fluctuations may be assigned to the
  photonic probe or the hadronic target.  Within this picture we
  attribute all of them to the target.}
\end{thebibliography}

\end{document}